\documentclass{aa}  
\usepackage{txfonts}
\usepackage{graphicx}
\usepackage{subfig}
\usepackage{epstopdf}
\usepackage{fixltx2e}
\usepackage{float}
\usepackage{natbib}

\newcommand{\CaII}{Ca~{\sc ii}}
\newcommand{\mets}{m~s\textsuperscript{-1}}

\begin{document}
\title{Evidence of photospheric vortex flows at supergranular junctions observed by FG/SOT (Hinode)}

\author{R. Attie \inst{1} 
\and D. E. Innes \inst{1}
\and H. E. Potts \inst{2}}

\institute{Max-Planck Institut f\"{u}r Sonnensystemforschung,
  37191 Katlenburg-Lindau, Germany
\and Department of Physics and Astronomy, University of Glasgow,
   Glasgow, G12 8QQ, UK}

\abstract
{Twisting motions of different nature are observed in several layers of the solar atmosphere. Chromospheric sunspot whorls and rotation of sunspots or even higher up in the lower corona sigmoids are examples of the large scale twisted topology of many solar features. Nevertheless, their occurrence at large scale in the quiet photosphere has not been investigated.}
{The present study reveals the existence of vortex flows located at the supergranular junctions of the quiet Sun.}
{ We use a 1-hour and a 5-hour time series of the granulation in Blue continuum and G-band images from FG/SOT to derive the photospheric flows. A feature tracking technique called Balltracking is performed to track the granules and reveal the underlying flow fields.}
{In both time series we identify long-lasting vortex flow located at supergranular junctions. The first vortex flow lasts at least 1 hour and is $\sim$20\arcsec~wide ($\sim$15.5~Mm). The second vortex flow lasts more than 2 hours and is $\sim$ 27\arcsec~wide ($\sim$21~Mm). }
{}


\maketitle

\section{Introduction}
Many measurements of the supergranular flows have been carried out in the era of MDI/Soho \citep{Shine00,Potts04,Potts08,Meunier08}. However, the time averaging and spatial smoothing necessary to derive the flows could not allow their analysis at large scale with both high dynamic and high resolution. The new capabilities of the Solar Optical Telescope (SOT) onboard Hinode allow systematic long and high resolution observations of the photospheric granulation. It is now possible to reveal more of the complexity of the underlying flows such as photospheric vortex flows which were first observed by \citet{Brandt88}. Our observations are described in section \ref{Observations} followed in section \ref{Analyses} by a summary of the tracking algorithm used to derive the flows. Section \ref{Results} presents the results of this analysis. Finally we discuss some scientific implications of these observations.

\section{Observations \label{Observations}}

We used two time series of de-rotated FG/SOT (Hinode) continuum images of the quiet Sun. They both have a field of view of $\sim$110\arcsec x110\arcsec ($\sim$86x86~Mm$^2$) :
\begin{itemize}
\item 1-hour time series of Blue Continuum and {\CaII} images on 10 April 2007 between 17:00 and 18:00 UTC with a cadence of 2~min at disk center.
\item 4-hours of contiguous G-Band and Blue continuum images on 07 Nov 2007 between 1:00 and 5:00 UTC near disk center at 3~min cadence. {\CaII} images were also taken during the Blue continuum time series at the same cadence. We use also the Na I stokes V of the Narrow Band Filter in FG/SOT for preliminary estimates of the line-of-sight topology of the magnetic field.
\end{itemize}

Both datasets were first binned onboard to 0.1\arcsec/px (binning~2x2) and corrected for dark currents, flat-fielding and instrumental jitter using standard calibration procedures. For better computational efficiency, we binned the calibrated images by another factor 2 giving a final resolution of 0.21\arcsec/px and an image size of 512 x 512 px$^2$.

\section{Analyses of photospheric flows \label{Analyses}}

\begin{figure}[h]
	\centering
	\includegraphics[trim=1cm 0.5cm 1cm 1.5cm, clip, width=0.35 \textwidth]{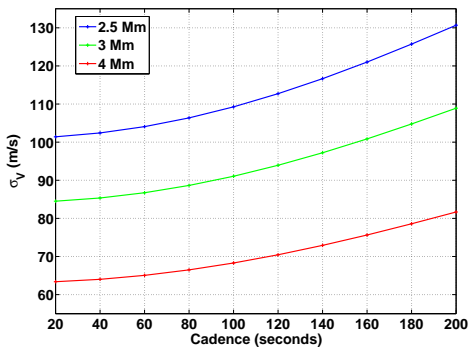}
	\caption{Uncertainty on the velocity measurements \(\sigma_v\) for a 45~minutes time averaging with respect to different instrumental cadences.\label{vel_errors1}}
	\end{figure}

\begin{figure}[h]
	\centering
	\includegraphics[trim=0.5cm 0cm 0.5cm 0.5cm, width=0.4\textwidth]{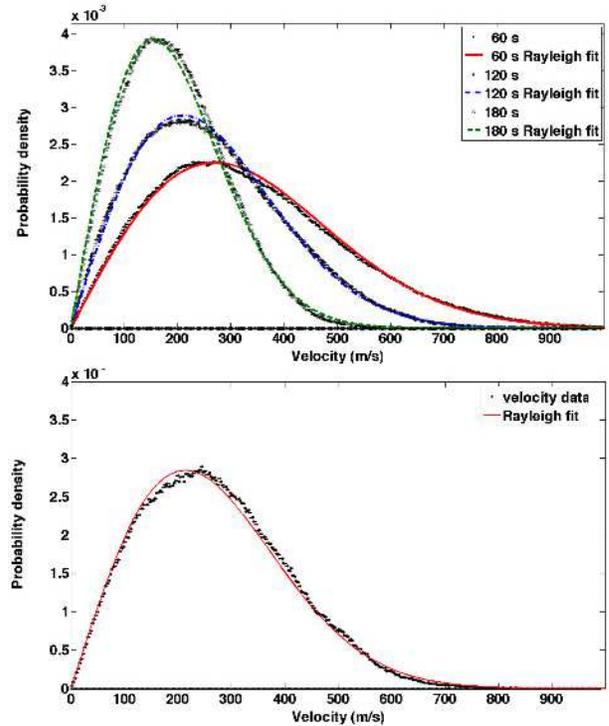}
	\caption{Top: Distributions of balltracked velocities at 3 cadences for test data on 11\textsuperscript{th} November 2007. Bottom: Distribution of the balltracked velocities for the 10\textsuperscript{th} April 2007 dataset. Time average is 30~min and spatial smoothing has a FWHM of 4~Mm. \label{distribution}}
	\end{figure}

To derive the photospheric flows from the continuum images, we used a tracking technique called Balltracking developed by \citet{Potts04}. In this section we present a short summary of this technique. \\
The balltracking technique considers the continuum images of the granulation in a three dimensional representation in which the intensity is seen as a geometrical height. Spherical tracers of a given mass and volume are released at a known position and settle in the local minima. In continuum images of the photospheric granulation these local minima are essentially constituted by the intergranular lanes. As the granules move, grow and shrink, the balls are pushed around and follow the resulting dynamic motion. The average width of the visible intergranular lanes is about 2 pixels (0.42\arcsec) in our datasets which is the radius we set for the tracking balls.
The outputs of the balltracking are velocity fields derived from the displacements of each tracking ball. They reflect the fast and stochastic small-scale granular motions. Therefore care is needed in the time averaging and spatial smoothing to reveal the underlying flows. 
Finally, a calibration procedure is needed to obtain physical units. As the balls take some time to settle in the local minima, the velocities are, on average, underestimated. The calibration consists in applying a known uniform velocity on subsets of the images to emulate an instrumental drift. The balltracking is then performed on the drifting subsets. Repeating the operation at different value of the drift velocity reveals that the calculated velocity has a linear response and requires a correction factor of $\sim$0.7. All smoothed balltracked velocities have been divided this factor.
\\From simulated velocity field and granular motions, \citet{Rieutord01} observed that granular tracking gives relevant underlying flow fields only at spatial scales larger than 2.5~Mm and at temporal scales of at least 30~min. Thus observations of the velocity fields below these values may not be meaningful. 
At these thresholds, \citet{Potts04} estimate that whatever the tracking algorithm used, for 1~min cadenced data and assuming perfect tracking of the granules, the uncertainty on the derived velocity flows would be $\sim$150~\mets. The uncertainty accounts for the finite scale of the granulation which makes the measurements spatially and temporally correlated. Thus the effective number of independent data points is smaller than the number of samples taken by the instrument.
Since our datasets have a cadence of 2 and 3~min, the uncertainty on the derived velocities have been recalculated for each,  following the formula in \citet{Potts04}:
\begin{equation}
 \sigma_v = \frac{\sigma_u}{\sqrt{n_r n_t}} 
 \end{equation}
 where \(n_r\) and \(n_t\) (respectively) are the effective spatial and temporal number of samples and \(\sigma_u\) is the standard deviation of a gaussian distribution approximating the velocity distributions of the granules. It is set to 800~\mets \citep{Roudier99}.
Our velocity fields were spatially smoothed over 4~Mm for both datasets. The results from the 1-hour and 5-hour observation series were averaged over 30~min and 45~min respectively. The choice of 4~Mm, which is above the 2.5~Mm threshold, is a compromise between a better accuracy of the velocity measurements and a spatial resolution small enough to distinguish the motions in narrow network lanes.
Figure \ref{vel_errors1} represents the uncertainty on any of the 2 components of the horizontal velocity flow with respect to different instrumental cadences for a 45~min average. The uncertainty on the velocity flows averaged over 45~min from the 3~min cadence dataset is then $\sim$78~\mets. At 2~min cadence and for a 30~min average (not shown here), the uncertainty is very similar. 
	\\Nevertheless, considering that the mean lifetime of the granules is $\sim$5~min, we believe that the granular motions cannot be properly sampled at cadences of 2 and 3~min. Moreover the removal of the instrumental jitter needs at least 2 images during the time over which the features do not change significantly. It cannot be guaranteed in these datasets. For these reasons, quantitative results may not be reliable. However, since we consider features of greater scale in time and space than the granular motions, the shape of the underlying velocity field can still be properly revealed \citep{Potts04}. As a test, we used a 3-hour-time-series with a cadence of 20~s from which 3 datasets with a cadence of 1,2 and 3~min were extracted. They were balltracked, smoothed and averaged over 4~Mm and 30~min. The upper pannel of figure \ref{distribution} shows that they are very well fitted by a Rayleigh distribution with a modal velocity decreasing with the cadence. Therefore we believe that lowering the cadence until 3~min acts as a low-frequency filter in the balltracking, which only picks up slower granules with lifetimes greater than 5~min. The bottom panel in figure \ref{distribution} represents the distribution of the velocities of the April dataset. The modal and mean velocities are between 200 and 300~\mets. These results are very close to the photospheric flows measured by \citet{Krijger02}. 
It is important to note that these values are smaller than the usual average speed measured in granular flows. A perfect tracking of the granular flow needs smaller smoothing and averaging. When we reduce these values, the modal and mean velocities indeed increase toward higher values more consistent with the results in \citet{Wang95}, \citet{Berger98} and \citet{Roudier99}. To track the underlying velocity flow, the smoothing and averaging must be increased to reduce the uncertainty introduced by the stochastic nature of the granular motions.

\section{Results \label{Results}}

\begin{figure*}
	\centering
	\subfloat[10\textsuperscript{th} Apr. 2007 18:15 UTC]{\label{FG_velvec1_SNAP_0001} \includegraphics[trim = 1cm 1cm 0cm 0.5cm, clip, width=0.5\textwidth]{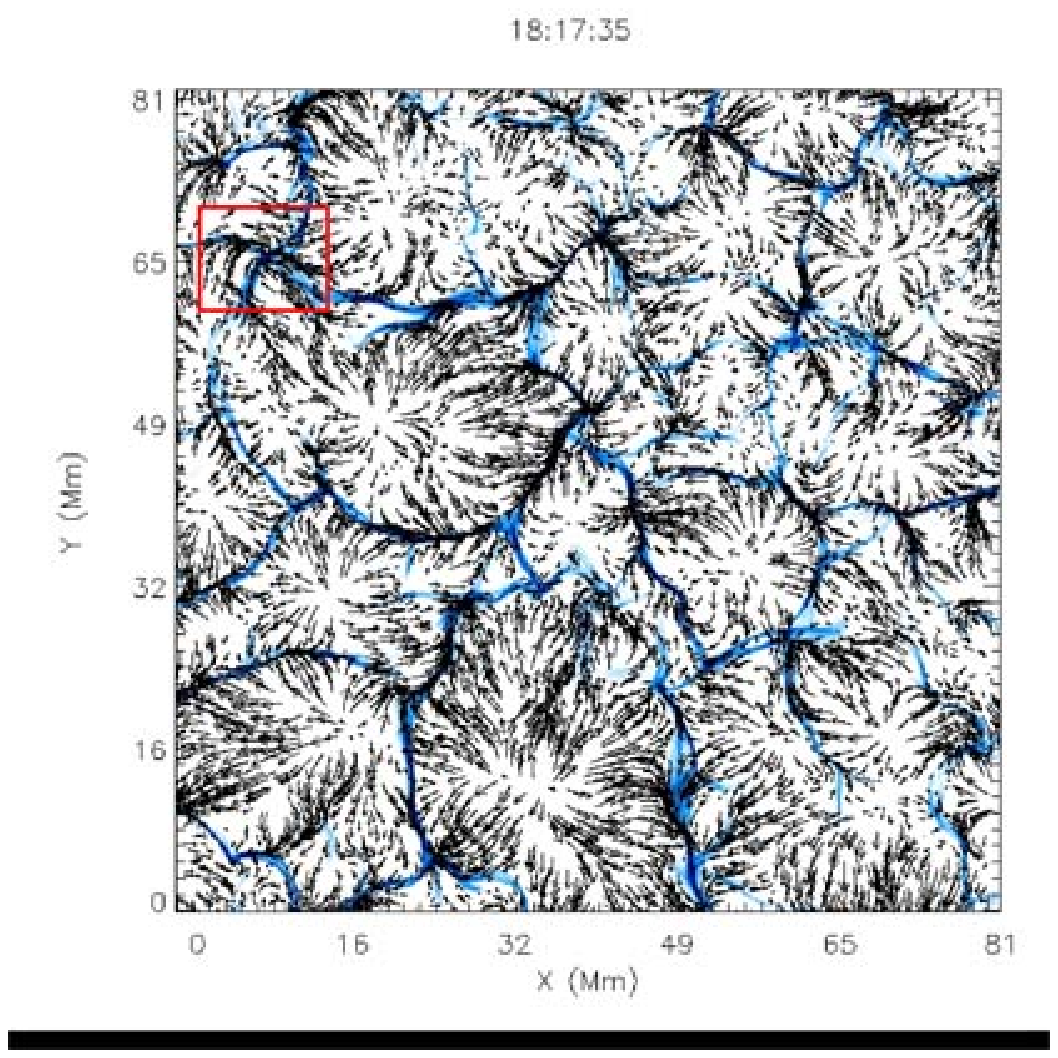}}
	\subfloat[10\textsuperscript{th} Apr. 2007 18:45 UTC]{\label{FG_velvec1_SNAP_0016} \includegraphics[trim = 1cm 1cm 0cm 0.5cm, clip, width=0.5\textwidth]{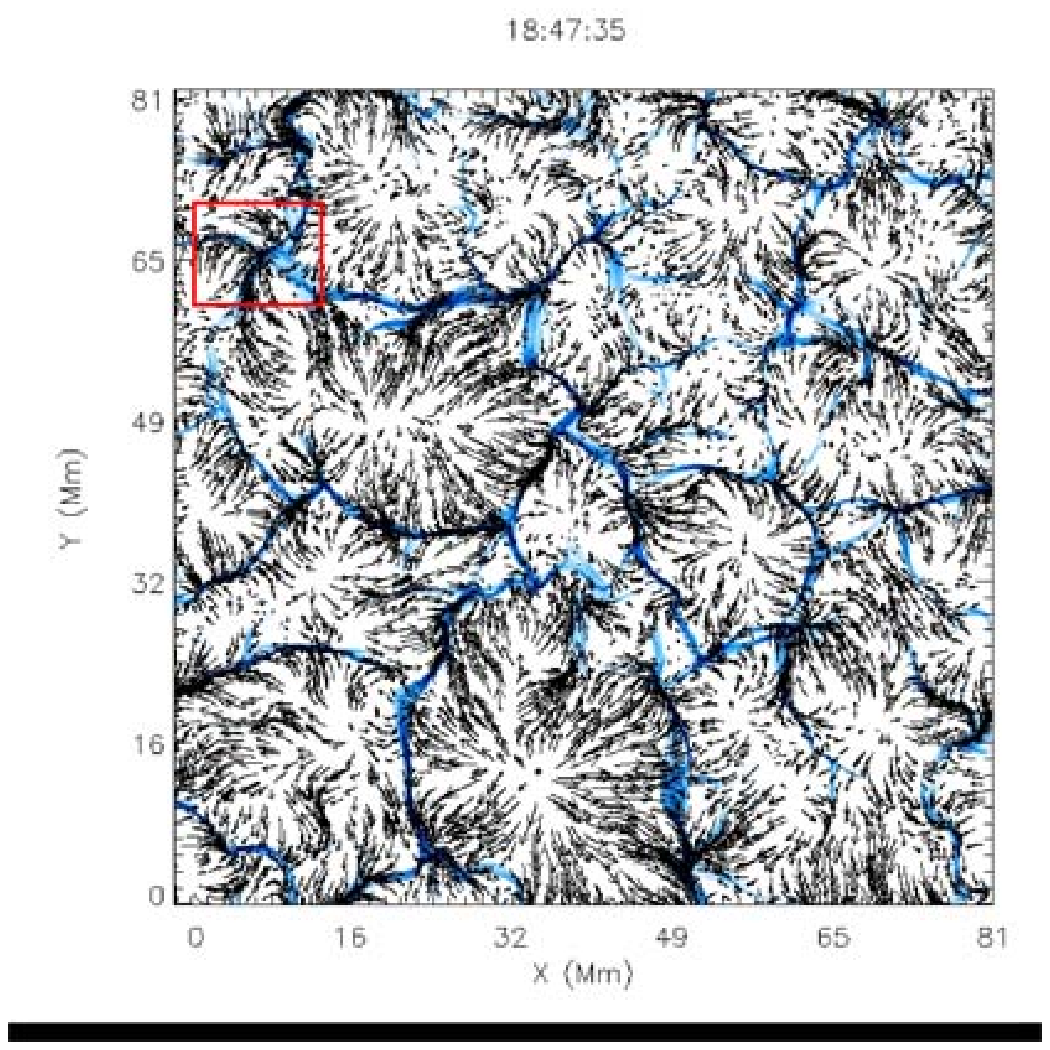}}
	\caption{Velocity vector fields of 10\textsuperscript{th} Apr. 2007 - FOV : $\sim$86x86~Mm$^2$. Red square : 15.5~Mm (diagonal) \label{velocity_1}}
	\end{figure*}

\begin{figure*}
	\centering
	\subfloat[07\textsuperscript{th} Nov. 2007 03:37 UTC]{\label{FG_velvec2_SNAP_0020} \includegraphics[trim = 1cm 1cm 0cm 0.5cm, clip, width=0.5 \textwidth]{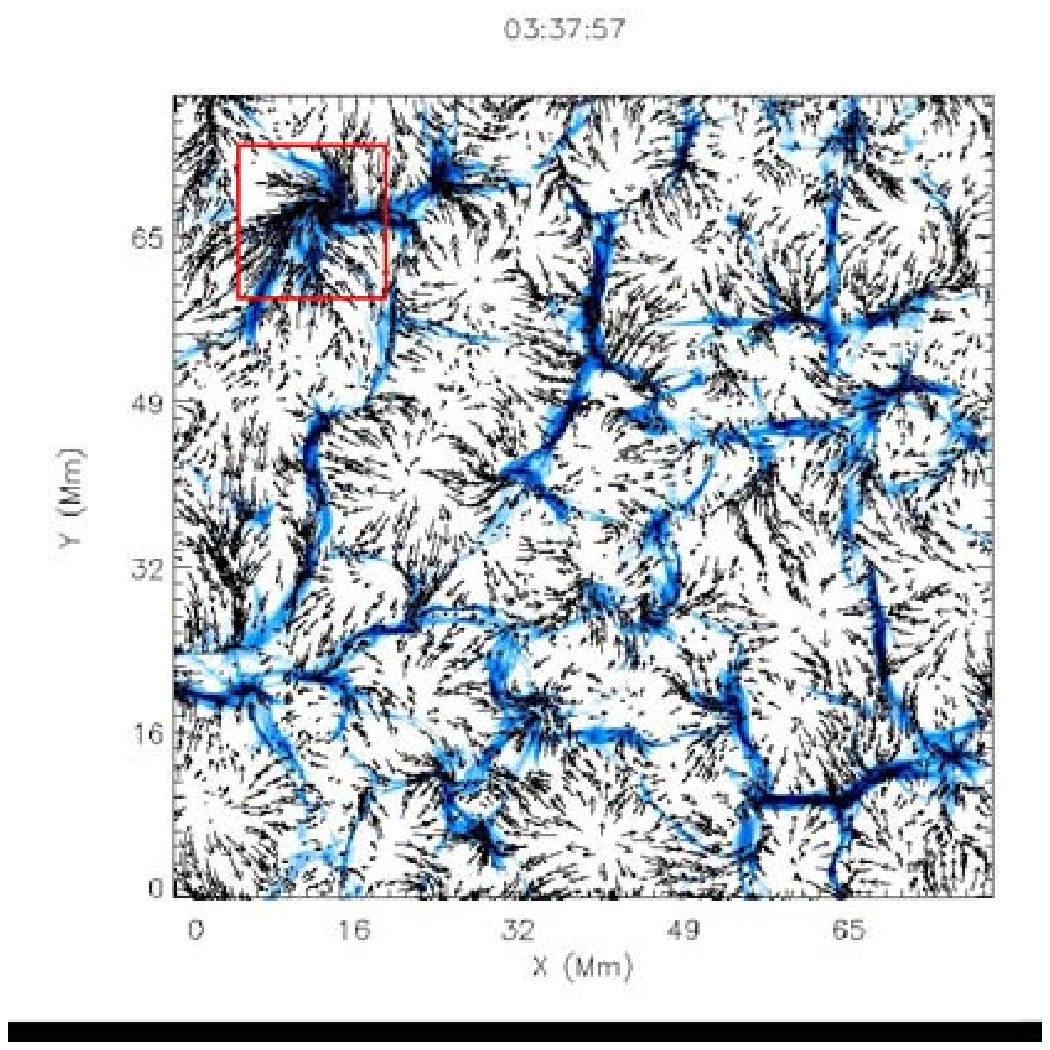}}
	\subfloat[07\textsuperscript{th} Nov. 2007 05:06 UTC]{\label{FG_velvec2_SNAP_0035} \includegraphics[trim = 1cm 1cm 0cm 0.5cm, clip, width=0.5 \textwidth]{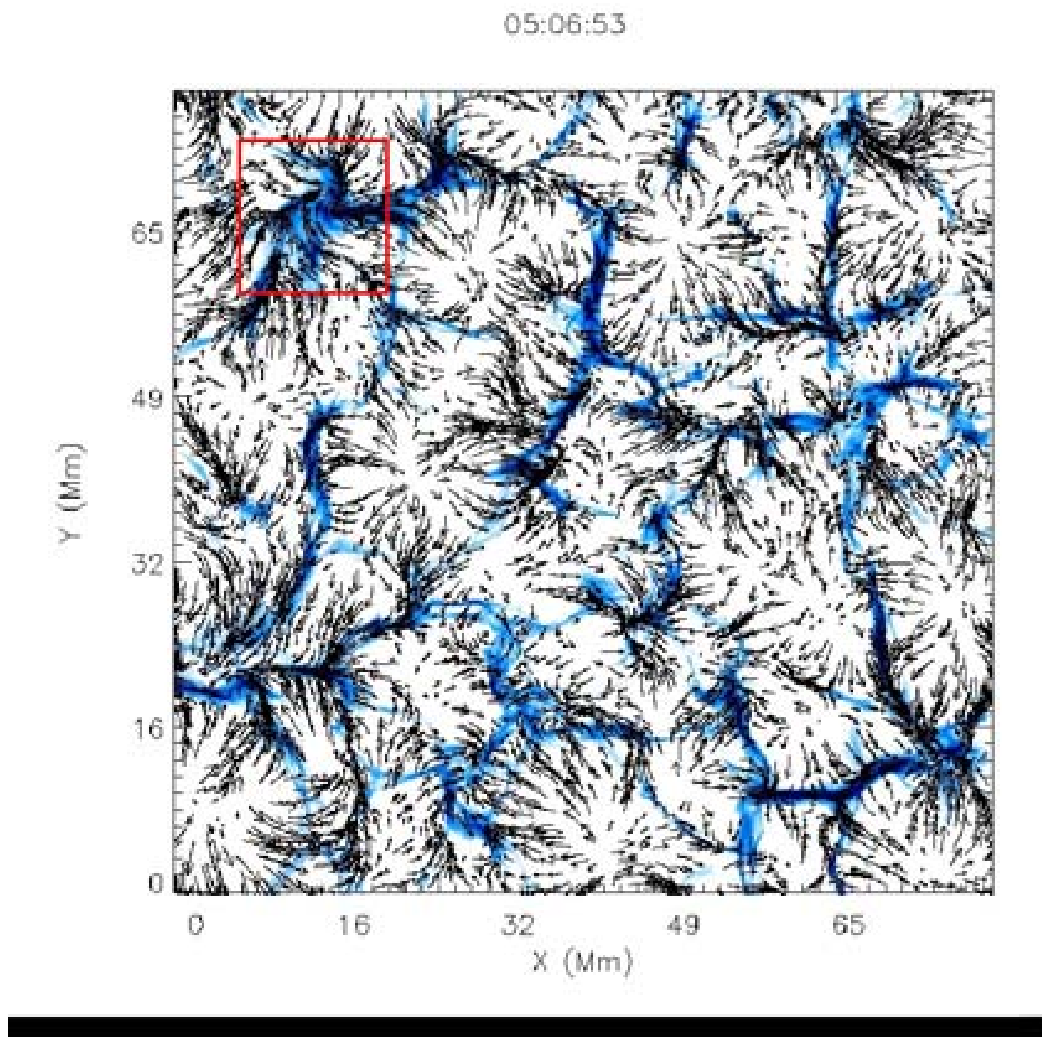}}
	\caption{Velocity vector field of 07\textsuperscript{th} Nov. 2007. The FOV is $\sim$86x86~Mm$^2$. Red square : 21~Mm (diagonal) \label{velocity_2}}
	\end{figure*}

\begin{figure}[h]
	\centering
	\includegraphics[width=0.4 \textwidth]{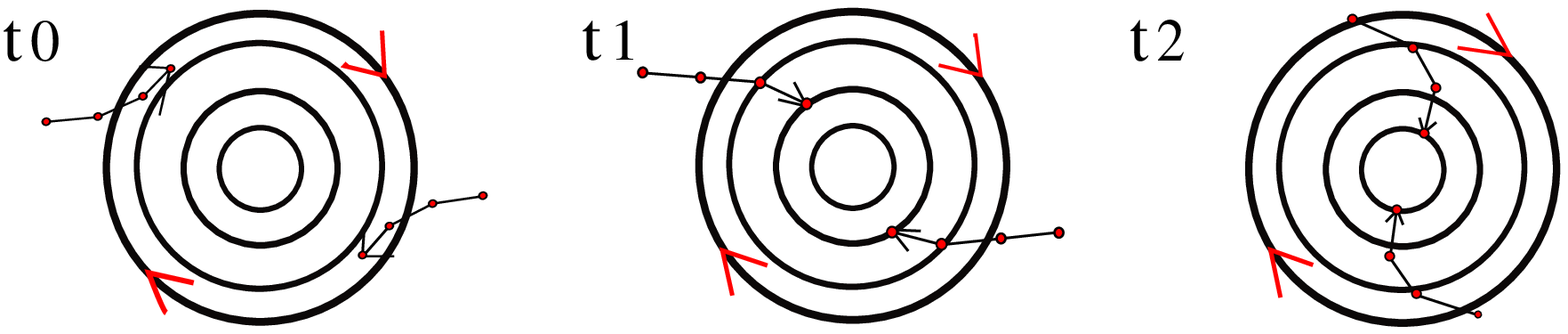}
	\caption{Sketch of the Lagrangian's representation of a vortex flow. The concentric circles represent the iso-contours of the tangential velocity increasing toward the center of the flow. The segmented arrows represent massless chained corks. \label{cartoon_vortex}}
	\end{figure}

The derived velocity flows from the balltracking on both datasets are shown in figure \ref{velocity_1} and \ref{velocity_2}. The upper (April) and bottom (November) row contain velocity fields at two different times. 
The black arrows are chained massless floating corks with an arrow head showing the direction of the flow (figure \ref{cartoon_vortex}). 
	
These objects are updated each time the corks are at a different position in the stream.
The light blue lanes in the background are the emphasized borders of the supergranular cells derived from an automated pattern recognition algorithm \citep{Potts08}. The representation is similar to what floating corks would show once they gather in the converging sites of the flows.
 A large scale converging vortex flow is clearly visible in each dataset. The first one is $\sim$20\arcsec ($\sim$15.5~Mm) and the second one is larger, $\sim$27\arcsec ($\sim$21~Mm). Both of them are located at supergranular junctions as indicated by the blue network lanes. There relative position in each field of view is very similar but is only a coincidence. However, these observations were made at similar latitudes and longitudes (near disk center). How it affects the occurrence of these vortex flows may be investigated with more observations and a greater statistic.
Figure \ref{cartoon_vortex} sketches how the chained corks are dragged towards the vortex flow.
	
\begin{figure}
	\centering
	\includegraphics[trim=2cm 1cm 2cm 1cm, clip, width=0.35 \textwidth, keepaspectratio=true]{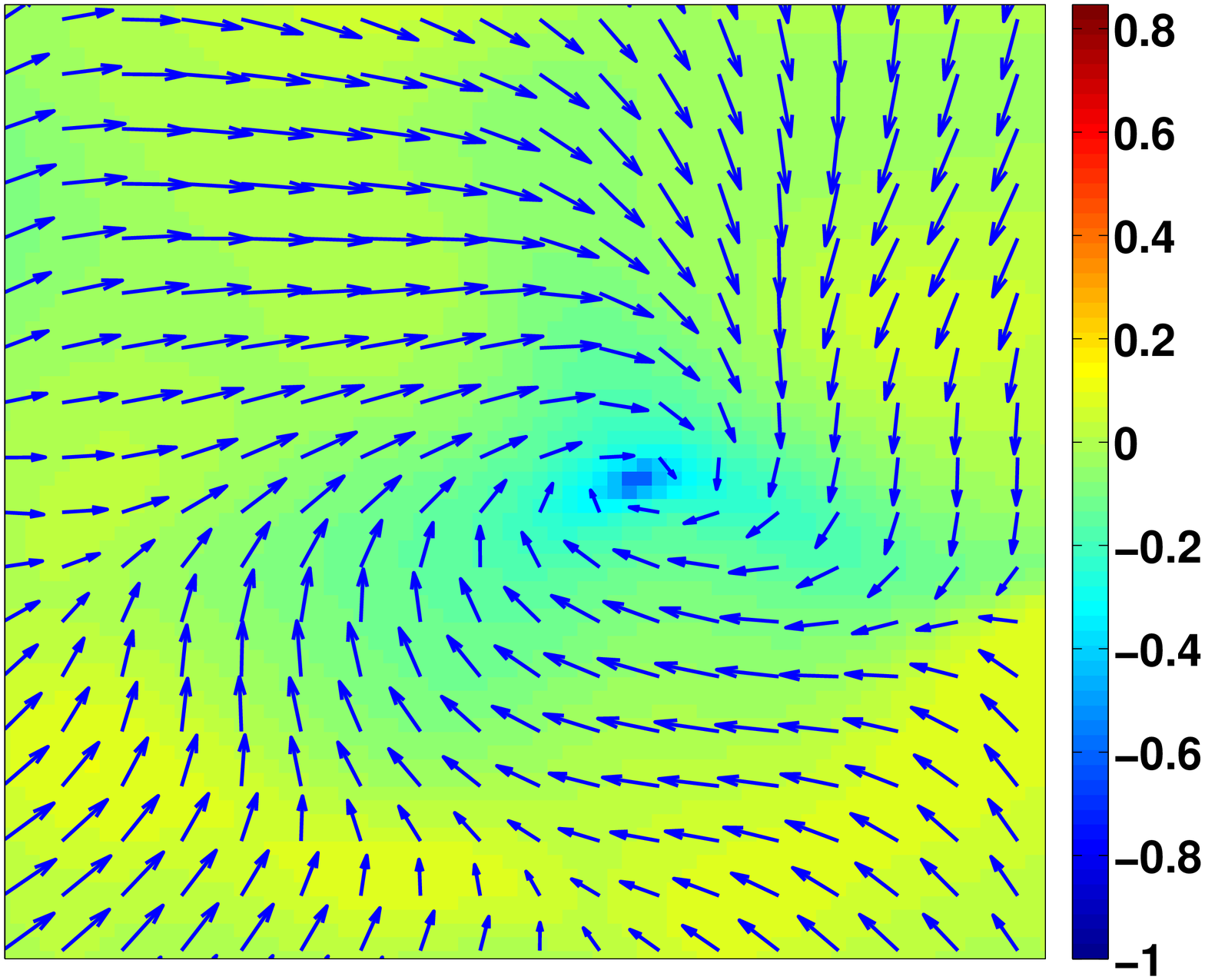}
	\caption{Eulerian representation of the flow in the red square of fig. \ref{velocity_2}. The colored background is the normalized vertical curl. \label{vortex}}
	\end{figure}

Figure \ref{vortex} is the Eulerian representation of the November velocity field centered on the vortex flow. The colored background is the normalized angular velocity. It is represented here as a visual indicator and should only be considered qualitatively (see section \ref{Observations}).
 We do not see the beginning of the vortex flow in the first dataset and it seems to continue longer than the 1 h time series. In the 2nd dataset, the onset seems to be between 2:30 and 3:10 and continue past 05:00, the end of the time series.
 		
\section{Discussion}
\begin{figure}
	\centering
	\includegraphics[trim= 1cm 1cm 0cm 0cm, clip, width=0.4\textwidth]{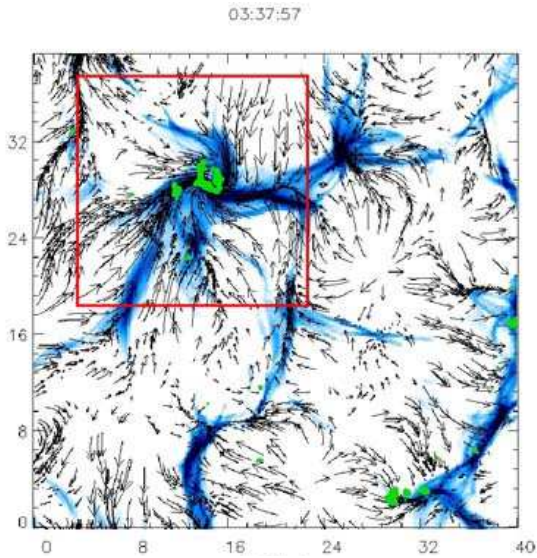}
	\caption{Zoomed-in velocity field of fig. \ref{velocity_2} with a filled contour of the maximum {\CaII} emission. The red square indicates the patch represented in fig. \ref{NaI} \label{vel_CaII}}
	\end{figure}

\begin{figure}
	\centering
	\includegraphics[width=0.5\textwidth]{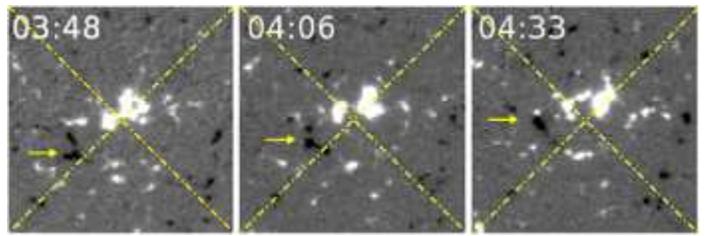}
	\caption{Na I stokes V (FG/SOT) from the November dataset at 3 different times centered on the vortex flows. The FOV of $\sim$20x20~Mm$^2$ is designated by the red square in fig. \ref{vel_CaII}. The yellow arrow points at a unipolar (black) magnetized fluid element. The dashed cross is a graphical fixed reference to help visualize the relative motions.\label{NaI}}
	\end{figure}

We demonstrate the occurrence of steady photospheric vortex flow located at supergranular junctions of the quiet Sun chromospheric network. Their influence extends to a radius of at least 7~Mm from the center of the vortex. The first observations of vortex flow was reported by \citet{Brandt88} and had a radius of 2.5~Mm. At granular scales, excess of vorticity has also been observed at the intersections of granular lanes \citep{Zirker93,Wang95}. Magnetic bright points following spirals has recently been reported by \citet{Bonet08}. Long-lived vortices are of primary interest because these are the ones in which mixed polarity magnetic fields can become entwined. In the quiet Sun, magnetic fields are swept by the supergranular flows to the boundaries and along the lanes to the core of the vortex, provoking flux cancellation and CME-like eruptions \citep{Innes08}. Figures \ref{vel_CaII} and \ref{NaI} illustrate the possible connection between the magnetic flux and the vortex flow. In figure \ref{NaI}, the black and white colors represent opposite polarities of the line of sight magnetic field. The positive (white) polarity near the image center coincides with the {\CaII} emission at the center of the vortex in figure \ref{vel_CaII}. In the time series, it is possible to see the negative polarity (black spot pointed by the yellow arrow) moving clockwise around the positive polarity.
\\Future observations must be done at higher cadence in order to reduce the uncertainty and to derive quantitative parameters such as the angular velocity and the magnetic helicity \citep{Welsch03}.
Observations from Hinode and in the near future from SDO could be used to compute the build-up of energy and relate it to activity in the chromosphere and the corona.
	
\bibliographystyle{aa}
\bibliography{aamnem99,sun}

\end{document}